\documentclass[aip,pof,reprint,a4paper]{revtex4-1}
\usepackage[ansinew]{inputenc} 
\usepackage[T1]{fontenc} 
\usepackage[english]{babel} 
\usepackage{amsmath} 
\usepackage{amsfonts}
\usepackage{amssymb}
\usepackage[pdftex,
			pdftitle={Comment},
            pdfauthor={Niklas Hietala},
            pdfsubject={},
            pdfkeywords={Kelvin wave, dispersion relation, LIA},
			bookmarks=true,            
            bookmarksopen=true,
            pdfpagemode = UseOutlines]{hyperref} 

\begin{document}

\title{Comment on ``Motion of a helical vortex filament in superfluid ${}^4$He under the extrinsic form of the local induction approximation''}
\author{Niklas Hietala}
\email[Electronic mail: ]{niklas.hietala@aalto.fi}
\author{Risto Hänninen}
\affiliation{O.V. Lounasmaa Laboratory, School of Science, Aalto University,  P.O. Box 15100, 00076 AALTO, Finland}
\date{\today}
\begin{abstract}
We comment on the paper by Van Gorder [``Motion of a helical vortex filament in superfluid ${}^4$He under the extrinsic form of the local induction approximation,'' Phys. Fluids \textbf{25}, 085101 (2013)].  We point out that the flow of the normal fluid component parallel to the vortex will often lead into the Donnelly--Glaberson instability, which will cause the amplification of the Kelvin wave. We explain why the comparison to local nonlinear equation is unreasonable, and remark that neglecting the motion in the $x$-direction is not reasonable for a Kelvin wave with an arbitrary wave length and amplitude. The correct equations in the general case are also derived.
\end{abstract}
\maketitle

Kelvin waves, which are helical perturbations of a vortex filament, are much studied in the context of superfluid turbulence due to the role they are believed to have in the dissipation of energy in the zero temperature limit. In his article\cite{gor13} Van Gorder studies a helical vortex at finite temperature and the normal fluid velocity being aligned along the same axis as the vortex. Van Gorder states that: ''for non-zero values of the superfluid friction parameters, there are no helical solutions studied for any of the models in the literature, full or approximate.'' We would argue that this is not the case. 

When the normal fluid flow is parallel to the vortex, solutions with a constant amplitude are only exceptions. In the presence of mutual friction helical solutions are usually decaying or unstable. It is well known that normal fluid velocity parallel to the vortex may lead to the Donnelly--Glaberson instability\cite{che73,gla74,ost75}. This instability causes the amplification of Kelvin waves. 
A simple derivation to show the Donnelly--Glaberson instability (in the small amplitude limit and when $\alpha' =0$) can be found for example in an article by Barenghi \emph{et al.}\cite{bar04} In his paper Van Gorder fails to mention the Donnelly--Glaberson instability completely.

Let us consider a straight vortex with a small amplitude Kelvin wave. If one is in the zero temperature limit, where there is no dissipation due to the mutual friction (the mutual friction parameters $\alpha$ and $\alpha'$ are zero), then all helical solutions stay as they are. If $\alpha \neq 0$ and we have a flow of the normal fluid component parallel to the vortex, then there is only one critical wave number for which a helical vortex keeps its shape: $k = U / \gamma$, where $k$ is the wave number, $U$ the velocity of the normal component and $\gamma=(\Gamma/4\pi) \ln(c/\langle \kappa \rangle a_0)$ ($\Gamma \approx 9.97 \times 10^{-4} \text{ cm}^2/\text{s}$ is the quantum of circulation, $c$ is a scaling factor of order of unity, $\langle \kappa \rangle$ is average curvature and $a_0 \approx 1.3 \times 10^{-8}$ cm is the effective vortex core radius)\cite{note}.
The condition for the Donnelly--Glaberson instability to occur is $k < U/\gamma$. In this wave number regime the Kelvin wave will grow exponentially. 
If $k > U / \gamma$, the wave will be damped leading eventually to a straight vortex. 
 
Mathematically a Kelvin wave with the critical wave number will remain as it is. Physically this kind of solution is not interesting, because small perturbations would cause it either to damp or grow. Also, for infinitely long vortex, with non-zero $U$, the instability always occurs at small enough $k$. The question of the stability of helical vortices even at the zero temperature is interesting. Already Betchov has shown that they are unstable\cite{bet65}. Umeki\cite{ume10} and Salman\cite{sal13} have shown additionally that under the local induction approximation (LIA) there is a breather solution of vortex motion, where the shape of the vortex filament is helical as $t \rightarrow \pm \infty$. Salman also showed that the breather excitations of the vortex can lead to self-reconnections and generation of small vortex rings.

The effects of the mutual friction on the dispersion relation of a Kelvin wave are also known\cite{bar85, don}. 
Mutual friction will shift the resonance frequency by a factor $(1- \alpha')$ and cause the resonance frequency to broaden (induce dissipation) by a factor proportional to $\alpha$. The presence of a normal fluid flow and a finite amplitude will further slightly modify the resonance frequency and damping, as will be shown below.

After this brief review of some known results, let us now have a look at Van Gorder's article. First of all, an important part of Van Gorder's article is to compare his results with approximations to the fully nonlinear form of the LIA. The approximations that were considered were a cubic model of Shivamoggi\cite{shi11} and local nonlinear equation (LNE) of Laurie \emph{et al.}\cite{lau10} Comparison to Shivamoggi's approximation is reasonable, since Van Gorder's and Shivamoggi's equations are obtained in a similar way. On the other hand, comparison to LNE is unreasonable. Van Gorder's equation includes mutual friction and LNE does not. This is very clear from the original paper by Laurie \emph{et al.}\cite{lau10} The local nonlinear equation was given in the context of studying the Kelvin wave cascade, which is believed to be essential for the energy dissipation in quantum turbulence at zero temperature. In the zero temperature limit there is no mutual friction.

When using the local induction approximation, the velocity of the vortex is
\begin{equation} \label{eq1}
\mathbf{v} = \gamma \kappa \mathbf{t} \times \mathbf{n} + \alpha \mathbf{t} \times ( \mathbf{U} - \gamma \kappa \mathbf{t} \times \mathbf{n}) - \alpha' \mathbf{t} \times ( \mathbf{t} \times (  \mathbf{U} - \gamma \kappa \mathbf{t} \times \mathbf{n})) \text{,}
\end{equation}
where $\kappa$ is the local curvature (not average curvature, as one might conclude from Van Gorder's paper), $\mathbf{t}$ and $\mathbf{n}$ are the unit tangent and normal vectors to the vortex filament and $\alpha$ and $\alpha'$ are the dimensionless friction parameters. Van Gorder assumes the normal fluid velocity to be $\mathbf{U} = U \mathbf{i}_x$ and says that then the equation of motion reduces to
\begin{equation} \label{schwartz}
\mathbf{v} = (1 - \alpha') \gamma \kappa \mathbf{t} \times \mathbf{n} + \alpha \mathbf{t} \times \mathbf{U} + \alpha \gamma \kappa \mathbf{n} - \alpha' U \mathbf{t} + \alpha' \mathbf{U} \text{.}
\end{equation}
However, the second last term should actually be $-\alpha' (\mathbf{t} \cdot \mathbf{U}) \mathbf{t}$. Van Gorder refers to a paper by Shivamoggi\cite{shi11} when he gives this equation. Shivamoggi assumed that the vortex is essentially aligned along the $x$-axis and that is why he apparently considered that $\mathbf{t} \cdot \mathbf{U} \approx U$. This assumption is not reasonable for a vortex with a Kelvin wave of arbitrary wave length and amplitude. The dot product should give an extra factor of $\frac{\text{d} x}{\text{d} s}$.

One should be able to discard the tangential term, because movement in the direction of the tangent doesn't change the form of the vortex, hence it is not important. However, the discard of this term would change the dispersion relation. This is not the only peculiarity that catches reader's eye. In Van Gorder's equation (2) the term  $- \alpha U \Phi_x$, corresponding to the component in the direction of $\mathbf{t} \times \mathbf{i}_x$, misses the factor $\frac{\text{d} x}{\text{d} s}$ coming from the tangent vector. 
The same mistakes appear also in other Van Gorder's articles\cite{gor12b, gor13b}.
It is also questionable, why $\omega$ is real in the equations (3) and (4), when it should be complex.
The imaginary part indicates the presence of the dissipation. 

Let us also consider, what is Van Gorder's dispersion relation of a Kelvin wave in the absence of the mutual friction. We get
\begin{equation} \label{vanDispRel}
\omega = \frac{\gamma k^2}{[1+(Ak)^2]^{3/2}} \text{.}
\end{equation}
In the small amplitude limit, $Ak \ll 1$ (where $x\approx s$, or $\frac{\text{d} x}{\text{d} s} \approx 1$), this correctly produces the known dispersion relation $\omega = \gamma k^2$. However, this differs from the known results for a large amplitude Kelvin wave, within the local induction approximation, which is given by\cite{son12}
\begin{equation} \label{large}
\omega = \frac{\gamma k^2}{\sqrt{1+(Ak)^2}} \text{.}
\end{equation}

In order to explain all these deviations from known results, let us see how Van Gorder has derived his equation. The very first assumption one must make, is to assume the vortex to be single-valued in $x$-direction. Then one can give the $y$- and $z$-coordinates as functions of $x$ and time: $y\equiv y(x,t)$ and $z\equiv z(x,t)$. For a Kelvin wave we have $\mathbf{r} = x\mathbf{i}_x + A \cos(kx-\omega t + x_0) \mathbf{i}_y + A \sin(kx-\omega t + x_0) \mathbf{i}_z$. Then from the fact $\text{d}s^2 = \text{d}x^2 + \text{d}y^2 + \text{d}z^2$ it also follows that
\begin{equation} \label{dxds}
\frac{\text{d}x}{\text{d}s} = \frac{1}{\sqrt{1 + y_x^2 + z_x^2}} \text{.}
\end{equation}
For a Kelvin wave this is $\frac{\text{d}x}{\text{d}s} =(1+ A^2k^2)^{-1/2}$. If $Ak \ll 1$, this could be approximated to be $\frac{\text{d}x}{\text{d}s} = 1- (1/2)(y_x^2+ z_x^2) = 1- (1/2)(Ak)^2$ (as Shivamoggi and van Heijst did\cite{shi10}), or even $\frac{\text{d}x}{\text{d}s}=1$, but Van Gorder makes no such approximation.

Because Van Gorder chooses to represent the vortex as $\mathbf{r} = x\mathbf{i}_x + y(x,t)\mathbf{i}_y + z(x,t)\mathbf{i}_z$, where $x\equiv x(s)$ is not a function of time, he has to assume that the velocity has no component in $x$-direction. This assumption is unreasonable. The motion of a helical vortex is purely rotational only when $Ak \rightarrow 0$. If $Ak \rightarrow \infty$, then a segment of the vortex is almost like a segment of a vortex ring, and the motion of the vortex consists of translation to $x$-direction. Usually the motion is a combination of both rotation and translation.

So, if Van Gorder assumes that $\mathbf{v} = y_t\mathbf{i}_y + z_t \mathbf{i}_z$, he should assume that $Ak \ll 1$. This should be also clear from the equations he gets, when setting
\begin{equation}
\frac{\text{d}\mathbf{r}}{\text{d}t} = \gamma \kappa \mathbf{t} \times \mathbf{n} \text{,}
\end{equation}
where $\mathbf{t} = \frac{\text{d} \mathbf{r}}{\text{d}s}$ and $\kappa \mathbf{n} = \frac{\text{d} \mathbf{t}}{\text{d}s}$. For simplicity, we consider the zero temperature case ($\alpha = 0$ and $\alpha' = 0$) here. With Van Gorder's assumption, $v_x =0$, this leads to three equations, namely:
\begin{eqnarray}
x_t &=& 0 = \gamma (y_x z_{xx} - z_x y_{xx})\left(\frac{\text{d}x}{\text{d}s}\right)^3 \text{, } \label{xt} \\
y_t &=& -\gamma z_{xx} \left(\frac{\text{d}x}{\text{d}s}\right)^3 \quad \text{and} \quad
z_t = \gamma y_{xx} \left(\frac{\text{d}x}{\text{d}s}\right)^3 \text{.} \label{yt zt}
\end{eqnarray}

The equations \eqref{yt zt} can be combined defining variable $\Phi = y + iz$, the derivative \eqref{dxds} is given by $\frac{\text{d} x}{\text{d}s} = (1 + |\Phi_x|^2)^{-1/2}$, and we get equation
\begin{equation} \label{eq:gor}
i \Phi_t + \gamma ( 1 + |\Phi_x|^2)^{-3/2} \Phi_{xx} = 0 \text{,}
\end{equation}
which gives the incorrect dispersion relation \eqref{vanDispRel}.
This is the equation which Van Gorder obtained and used in his earlier works\cite{gor12,gor12c,gor12b}. In most of his papers Van Gorder doesn't mention the constraint resulting from $x_t=0$ (equation \ref{xt}), nor does Shivamoggi and van Heijst\cite{shi10,shi11} or Umeki\cite{ume13} ever consider this. Only in his first article\cite{gor12} concerning LIA in extrinsic coordinate space Van Gorder does mention this. 

The condition $y_x z_{xx} - z_x y_{xx} = 0$ can be given with the help of $\Phi$ in the following way: $\Phi_x^* \Phi_{xx} - \Phi_x \Phi_{xx}^* =0$ (in other words the Wronskian of $\Phi_x$ and $\Phi_x^*$ is zero). For a Kelvin wave,  $\Phi=A \exp[i(kx-\omega t + x_0)]$, the equation \eqref{xt} gives $\gamma A^2k^3(\frac{\text{d}x}{\text{d}s})^3 = 0$. Here we see that one has to assume that $Ak \ll 1$ in order to neglect the term on the left hand side.

In the paper under the consideration, it seems that in making the change of variable for the velocity term in the direction of the normal vector, he exploits this requirement. However, this is not totally clear, since if the mutual friction terms are included, then the expression for $x_t$ is more complicated. Typos in the equation giving the normal vector $\kappa \mathbf{n}$ may cause even more confusion (same typos appear also in Ref. \citenum{gor12b} and \citenum{gor13b}). The normal vector should be
\begin{eqnarray} \label{normal}
\kappa \mathbf{n} &=& -(y_x y_{xx} + z_x z_{xx})\left(\frac{\text{d}x}{\text{d}s}\right)^4\mathbf{i}_x 
+ \left( y_{xx}\left(\frac{\text{d}x}{\text{d}s}\right)^2 - y_x (y_x y_{xx} + z_x z_{xx})\left(\frac{\text{d}x}{\text{d}s}\right)^4\right)\mathbf{i}_y \nonumber \\
&& + \left( z_{xx}\left(\frac{\text{d}x}{\text{d}s}\right)^2 - z_x (y_x y_{xx} + z_x z_{xx})\left(\frac{\text{d}x}{\text{d}s}\right)^4\right)\mathbf{i}_z 
\\
&=& -(y_x y_{xx} + z_x z_{xx})\left(\frac{\text{d}x}{\text{d}s}\right)^4\mathbf{i}_x  + \left(y_{xx} + z_x ( z_x y_{xx} - y_x z_{xx}) \right) \left(\frac{\text{d}x}{\text{d}s}\right)^4\mathbf{i}_y \nonumber \\
&& + \left(z_{xx} - y_x ( z_x y_{xx} - y_x z_{xx}) \right) \left(\frac{\text{d}x}{\text{d}s}\right)^4\mathbf{i}_z \text{.} \nonumber
\end{eqnarray}

If Van Gorder would simply neglect all terms with $A^2k^2$, he would recover all the known results obtained in the small amplitude limit. 
The amplitude of a helical wave wouldn't be anymore related to the wave number. One would obtain the same critical wave number $k = U / \gamma$ as with Shivamoggi's model\cite{shi10} and which was shown also in the article by Barenghi \emph{et al.}\cite{bar04}

Still, one would be left wondering why the term coming from the tangential velocity shows in the dispersion relation. One should be able to discard the tangential velocity components. This problem is solved if one takes into account the movement in the $x$-direction. Having $x \equiv x(s,t)$, and then $\mathbf{v} = \frac{\text{d} \mathbf{r}}{\text{d} t} = x_t \mathbf{i}_x + (y_t + y_x x_t) \mathbf{i}_y + (z_t + z_x x_t) \mathbf{i}_z$, will give $\Phi_t = y_t + iz_t = 0$, if one has $\mathbf{v} = K \mathbf{t} = K \frac{\text{d} x}{\text{d} s}( \mathbf{i}_x +y_x \mathbf{i}_y + z_x \mathbf{i}_z)$, where $K$ is an arbitrary constant. In other words, terms parallel to $\mathbf{t}$ have no effect on $\Phi_t$.

Using $x \equiv x(s,t)$, the equation
\eqref{schwartz}
can be expressed in the potential form as
\begin{eqnarray} \label{full}
&&i \Phi_t 
- i \alpha \gamma \left(\frac{\text{d} x}{\text{d} s}\right)^4 (\Phi_{xx} + \Phi_x \Phi_x^* \Phi_{xx})
+ i \alpha' U \Phi_x 
- \alpha U  \left(\frac{\text{d} x}{\text{d} s}\right) \Phi_x \nonumber \\
&& + (1-\alpha')\gamma \left(\frac{\text{d} x}{\text{d} s}\right)^3 \left( \Phi_{xx} + \frac{1}{2} \Phi_x (\Phi_x^* \Phi_{xx} - \Phi_x \Phi_{xx}^* ) \right)
= 0
 \text{.}
\end{eqnarray}
The resulting dispersion relation in zero temperature is the dispersion relation for a large amplitude Kelvin wave \eqref{large}. The dispersion relation with mutual friction is
\begin{eqnarray}
\text{Re}( \omega ) =& (1-\alpha') \dfrac{\gamma k^2}{\sqrt{1 + (Ak)^2}} + \alpha' U k &= (1-\alpha') \omega_0 + \alpha' Uk \\
\text{Im}( \omega ) =& \dfrac{\alpha U k}{\sqrt{1 + (Ak)^2}} - \dfrac{\alpha \gamma k^2}{1 + (Ak)^2} &= \alpha (1+ A^2k^2)^{-1/2} (Uk- \omega_0) \text{,} 
\end{eqnarray}
where $\omega_0 = \gamma k^2 (1+A^2k^2)^{-1/2}$ is the zero temperature dispersion relation.

Requiring $\text{Im}(\omega) =0$, will give us the critical wave number for the Donnelly--Glaberson instability, i.e. the wave number for which the amplitude stays constant in time. We will notice that the critical $k$ depends actually on the amplitude. We get either $k=0$ (a straight vortex) or
\begin{equation} \label{critical}
k = \frac{U}{\sqrt{\gamma^2 - U^2 A^2}} \text{.}
\end{equation}
In the small amplitude limit this is the familiar $k = U / \gamma$. 
From the denominator we get an additional condition
\begin{equation} \label{cond}
\gamma > UA \text{.}
\end{equation}
This is a reasonable condition. If one had $\gamma < UA$, then a vortex ring with radius $A$ moving in the direction of the flow of the normal component would grow.

In the zero temperature limit the equation \eqref{full} reduces to
\begin{equation} \label{full0}
i \Phi_t + \gamma \left(\frac{\text{d}x}{\text{d}s}\right)^3 [ \Phi_{xx} + \frac{1}{2} \Phi_x (\Phi_x^* \Phi_{xx} - \Phi_x \Phi_{xx}^*)] = 0 \text{.}
\end{equation}
An easier way to derive equation (\ref{full0}) would have been using the Hamiltonian formulation. The Hamiltonian corresponding to LIA\cite{svi95} and the equation of the motion are given by
\begin{eqnarray}
H[\Phi] &=& 2 \gamma \int \text{d}x \sqrt{1+|\Phi_x|^2} \\
i \Phi_t &=& \frac{\delta H[\Phi]}{\delta \Phi^*} \text{.}
\end{eqnarray}
The equation (\ref{full0}) is the same as for example equation (4) in article by Boffetta \emph{et al.}\cite{bof09} The advantage of deriving the equation via formulating LIA in the Cartesian coordinates is that the inclusion of the mutual friction is straightforward.
In their paper Boffetta \emph{et al.} have also an equation of motion with a mutual friction term (equation 27), where this term is proportional to $\Phi$, when it evidently should be proportional to $\Phi_{xx}$.

\begin{acknowledgments}
The authors would like to thank H. Salman for useful discussions on this comment. 
This work is supported by the Academy of Finland (Grant No. 218211 and 263798)
and in part by the European Union 7th Framework Programme (FP7/2007-2013, Grant No. 228464 Microkelvin)
and by the Academy of Finland through its LTQ CoE grant (Project No. 250280).
\end{acknowledgments}


\begin{thebibliography}{19}
\bibitem{note} In our notation we have included the factor $1/4 \pi$ in $\gamma$ and kept the units, since Van Gorder's notation is only partly dimensionless due to the units in $a_0$. This has no effect on any of the equations below.
\end{thebibliography}
\end{document}